\documentclass[english]{article}
\usepackage[T1]{fontenc}
\usepackage[latin9]{inputenc}

\usepackage{amsmath,graphicx} 

\makeatletter

\providecommand{\tabularnewline}{\\}

\makeatother

\usepackage{babel}
\begin{document}

\title{Spectral Madness - Widening the Scope of a Partial Dynamic Symmetry}

\author{Wesley Pereira, Ricardo Garcia Larry Zamick and Alberto Escuderos\\
Department of Physics and Astronomy, Rutgers University, Piscataway,
New Jersey 08854\\}

\maketitle
\begin{abstract}
If one examines two-body matrix elements from experiment, one notices
that not only J=0 T=1 lies low, but also J=1 T=0 and J=J$_{max}$
=2j T=0. It is sometimes thought that one needs both T=1 and T=0 two-body
matrix elements to get equally spaced spectra of even I states, i.e.,
vibrational spectra. We here attempt to get equally spaced levels
with only those that have T=1 (even J). As an example, we perform
single-j calculations (f$_{7/2}$ ) in $^{44}$Ti and $^{46}$Ti.
We then shift gears and decide to play around with the input two particle
matrix elements (not worrying about experiment) to generate interesting
spectra, e.g., rotational spectra, with and then without T=0 two-body
matrix elements. We also consider simple interactions (e.g. \textquotedbl{}123\textquotedbl{},''1234''
and ``12345'') and find an expanded partial dynamical symmetry.
\end{abstract}

\section{Introduction}

In a Nature article by B. Cederwall et al.{[}1{]}, they report findings
of equally spaced levels J=0, 2, 4, 6 in the 8 hole nucleus $^{92}$Pd.
Their calculated B(E2)'s {[}1.2{]} are not consistent with a simple
vibrational interpretation. In the ``supplements material'' of the
Nature article they feel that they have an isoscalar spin aligned
coupling scheme and emphasize configurations in which a neutron and
proton couple to J=J$_{max}$=9 in the g$_{9/2}$ shell. Cleary, they
emphasize the importance of odd-J T=0 two-body matrix elements. Robinson
et al.{[}3{]} noted that in a large space calculation the static quadrupole
moment of the lowest 2$^{+}$ state of $^{92}$Pd was very small,
consistent with the vibrational picture. On the other hand, $^{96}$Cd
turned out to be prolate, and $^{88}$Ru, oblate.

Previous to this, Robinson et al.{[}4{]} made a study of doing full
fp calculations of another 8 particle system, $^{48}$Cr. They compared
the results of including all two-body matrix elements with those in
which all T=0 two-body matrix elements were set to zero. The qualitative
discussions in the 2 cases were quite different. In a lanl preprint
by Robinson et al.{[}5{]}, a figure is shown which confirms the implications
of refs {[}1,2{]} that dire consequences occur when T=0 two-body matrix
elements are set to zero. With one of the interactions used, the first
6$^{+}$ and 8$^{+}$ states are nearly degenerate and the B(E2)'s
8 to 6 and 6 to 4 are very small. On the other hand, for $^{48}$Cr
the emphasis was that it was hard to tell which of the calculations
agreed better with experiment -- full interaction of T=1 only. Sure,
there were some differences, but both spectra looked similar -- sort
of vibrational, but with a tendency to rotation. The B(E2)'s were
larger in the full calculation, but this could be accomodated to a
large extent by changing the effective charges.

Although the above works serve as a stimulus for we are about to do,
we would like to separate ourselves from discussions and arguments
about the relative merits therein. Rather, we start anew and address
the problem of how two-body matrix elements affect the spectra of
more complex nuclei.

\section{Attempt to generate near equally spaced spectra with only T=1 two-body
matrix elements -$^{44}$Ti}

In 1963 and 1964, calculations were performed to obtain wave functions
and energy levels in the f$_{7/2}$ shell by Bayman et al. {[}6{]},
McCullen et al. {[}7{]} and French et al. {[}8{]}. At that time, the
T=1 two-body matrix elements were well known but not so T=0. In 1985,
the T=0 matrix elements were better known and the calculations were
repeated by Escuderos et al. {[}9{]}. The two-particle matrix elements,
obtained mainly from the spectrum of $^{42}$Sc, are shown in the
table. Note that not only J=0 T=1, but also J=1 T=0 and J=7 T=0, are
low lying .

We next make a search for a set of T=1 two-body matrix elements which
will give close to equal spaces spectra for even I states in $^{44}$Ti,
and for which all the T=0 (odd-J) matrix elements are set equal to
zero. We find a good choice for J=0, 2, 4, 6 are, respectively, 0.00,
1.00, 1.60, 2.00.

\begin{center}
\textbf{Table I: Spectra of $^{42}$Sc} 
\par\end{center}

\begin{center}
\begin{tabular}{|c|c|c|c|c|c|}
\hline 
J  & MBZE  & T=0  & MBZE  & T=1 only  & MeV\tabularnewline
\hline 
\hline 
0  & 0.0000  & 1  & 0.6111  & 0  & 0.0000\tabularnewline
\hline 
2  & 1.5865  & 3  & 1.4904  & 2  & 1.0000\tabularnewline
\hline 
4  & 2.8153  & 5  & 1.5101  & 4  & 1.6000\tabularnewline
\hline 
6  & 3.2420  & 7  & 0.6163  & 6  & 2.0000\tabularnewline
\hline 
\end{tabular}
\par\end{center}

\section{Spectra (MeV) of $^{44}$Ti and $^{46}$Ti}

\begin{center}
\textbf{Table II: Spectra of $^{44}$Ti and $^{46}$Ti} 
\par\end{center}

\begin{center}
\begin{tabular}{|c|c|c|c|c|}
\hline 
I  & MBZE $^{44}$Ti  & MBZE $^{46}$Ti  & T=1 only $^{44}$T  & T=1 only $^{46}$Ti\tabularnewline
\hline 
\hline 
0  & 0.0000  & 0.0000  & 0.0000  & 0.0000\tabularnewline
\hline 
2  & 1.1631  & 1.1483  & 0.8302  & 0.8397\tabularnewline
\hline 
4  & 2.7900  & 2.2225  & 1.5723  & 1.5535\tabularnewline
\hline 
6  & 4.0618  & 3.1575  & 2.1508  & 1.9492\tabularnewline
\hline 
8  & 6.0842  & 4.8720  & 3.4643  & 3.1042\tabularnewline
\hline 
10  & 7.3839  & 6.3334  & 4.2524  & 4.0223\tabularnewline
\hline 
12  & 7.7022  & 8.0257  & 4.8524  & 4.9490\tabularnewline
\hline 
\end{tabular}
\par\end{center}

We have not made an exhaustive search the optimum T=1 matrix elements
so as to obtain an equally spaced spectrum of $^{44}$Ti, but it is
more than sufficient to get the point across. With a perfect vibrator
the I=12$^{+}$ state would be at 6$\times$0.8302=4.9812 MeV. In
fact, it is at 4.8524 MeV. The 8 $\rightarrow$ 6 splitting is significantly
dfferent for the 2 $\rightarrow$ 0 splitting of 0.8397 MeV. It is,
in fact, 1.3105 MeV.

\begin{center} 
  \includegraphics[width=.5\textwidth]{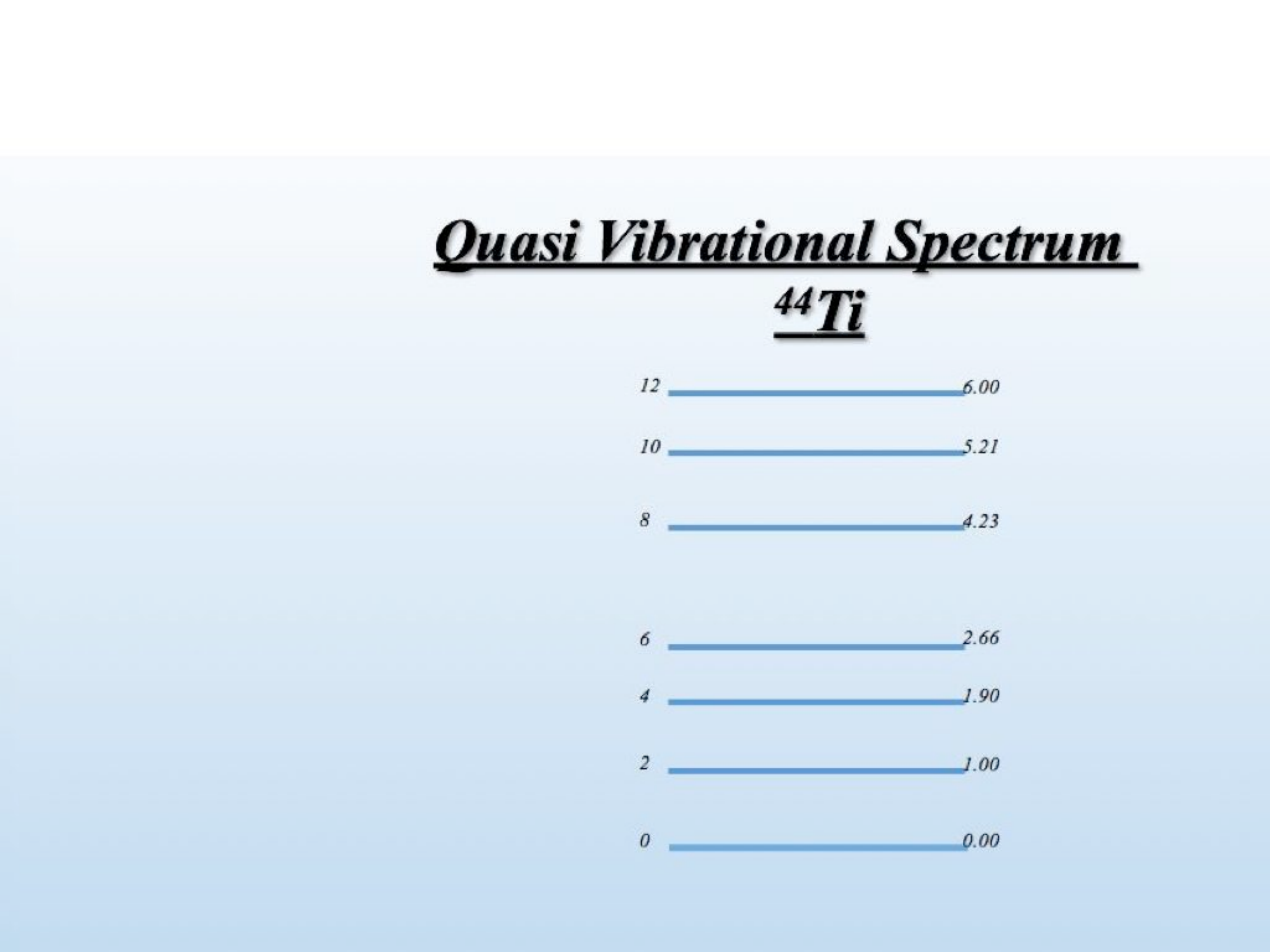}%
\includegraphics[width=.5\textwidth]{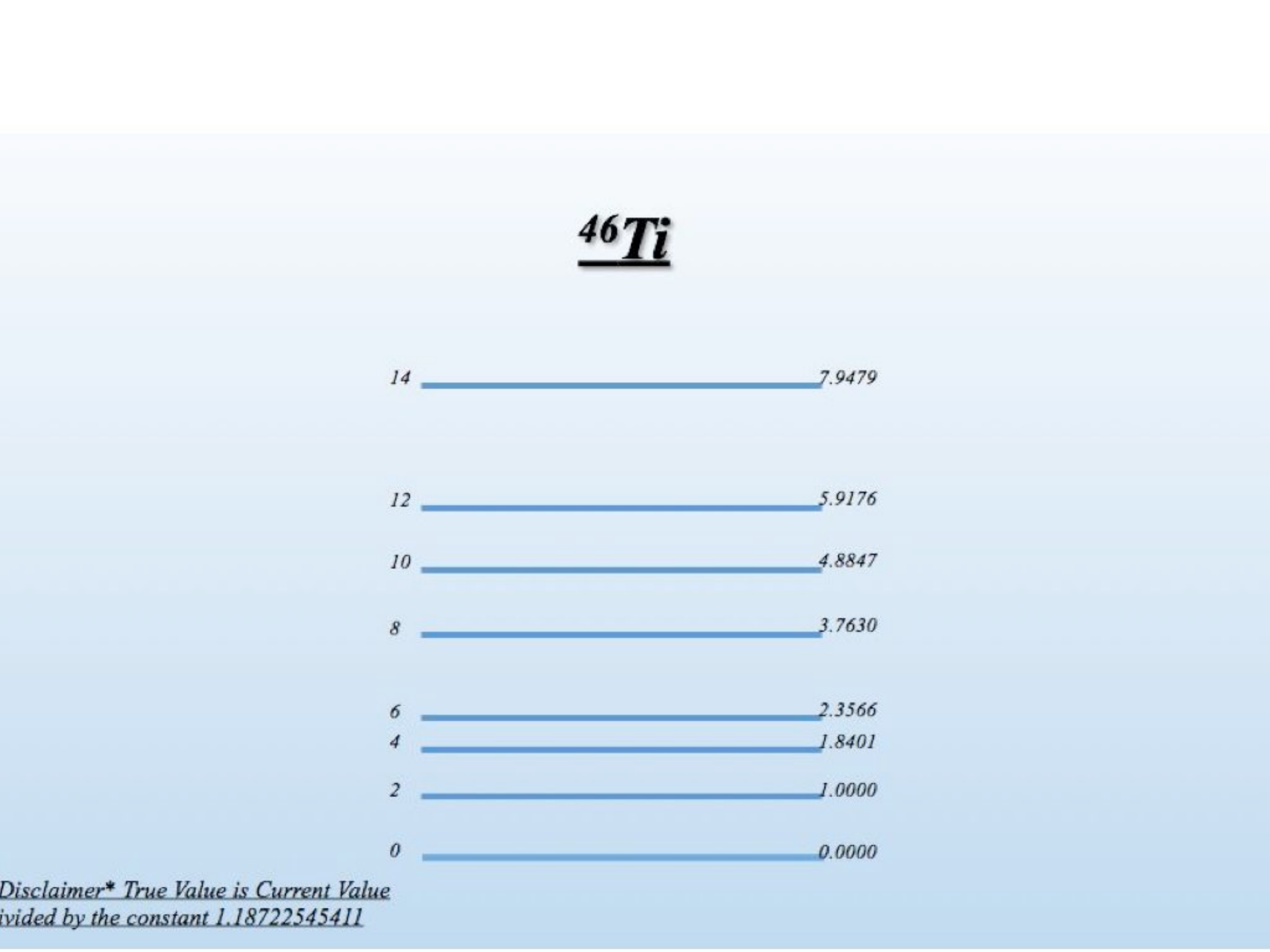}
\end{center}

\section{Rotational Spectra.}

We can also play the game of obtaining rotational spectra. This is
easier. We consider 3 cases:

1. Set all two-particle matrix elements to J(J+1). Then it is easy
to show that for $^{44}$Ti, one also gets a perfect I(I+1) spectrum
with the I=2 state at 6 MeV and the I=12 at 156 MeV.

\begin{center}
\textbf{Table III: Perfect rotator spectrum for $^{44}$Ti} 
\par\end{center}

\begin{center}
\begin{tabular}{|c|c|c|}
\hline 
I  & E  & T=1 only scaled\tabularnewline
\hline 
\hline 
0  & 0  & 0\tabularnewline
\hline 
2  & 6  & 6\tabularnewline
\hline 
4  & 20  & 19.58\tabularnewline
\hline 
6  & 42  & 41.16\tabularnewline
\hline 
8  & 72  & 70.16\tabularnewline
\hline 
10  & 110  & 107.80\tabularnewline
\hline 
12  & 156  & 150.61\tabularnewline
\hline 
\end{tabular}
\par\end{center}

2. We next refer to Table III. In the first spectral column we have
the perfect rotator spectrum. We obtain this from a J(J+1) 2-body
spectrum where both T=0 and T=1 2-body matrix elements are included.
In the next column we have a spectrum in which only the T=1 matrix
elements are included but we rescale them so the first 2$^{+}$ state
comes at 6 MeV, the same as in the first column. Without rescaling
the 2$^{+}$ state would be at 4.625 MeV and the 12$^{+}$ at 116.10
MeV. We do not obtain a perfest rotational spectrum but still a fairly
good one.

3. If we set all two-body T=1 matrix elements to zero and set the
T=0 ones to J(J+1), we get a most fascinating spectrum. The even I
states with I=0, 2, 4, 6 and 8 are at zero energy whilst the 10$^{+}$
and 12$^{+}$ states are at 94 MeV and 156 MeV respectively. While
at first glance surprising, there is an easy explanaton for the multiple
ground state zeros. All states of zero energy have isospin T=2. They
are therefore double analogs of states in$^{^{44}}$Ca, which in our
model space consists of 4 valence neutrons. The neutron-neutron interacton
occurs only in the T=1 channel, but in this case all three of the
T=1 matrix elemnts are set to zero. This explains why the T=2 states
in $^{44}$Ti lie at zero energy. A closer examination shows that
there are two I=2$^{+}$ and two I=4$^{+}$ states at zero energy.
But it is well known that in $^{44}$Ca such states occur twice. One
I=2$^{+}$ state has seniority v=2 and the other v=4; likewise for
I=4$^{+}$. Note that there are no I=10$^{+}$ or 12$^{+}$ states
of the (f$_{7/2}$)$^{4}$ configuration in $^{44}$Ca. This explains
why these states do not appear at zero energy in $^{44}$Ti.

In summary, with a J(J+1) two-body interaction one gets a perfect
rotations spectrum in more complex nuclei, e.g., $^{44}$Ti. If this
interaction is only in the T=1 channel, one gets an imperfect, but
still fairly good rotational spectrum. If this interacton is only
in the T=0 channel one gets a spectrum with multiple T= 2 degeracies
at zero energy and the spectum does not at all resemble a rotational
spectrum.

\begin{center} 
  \includegraphics[width=\textwidth]{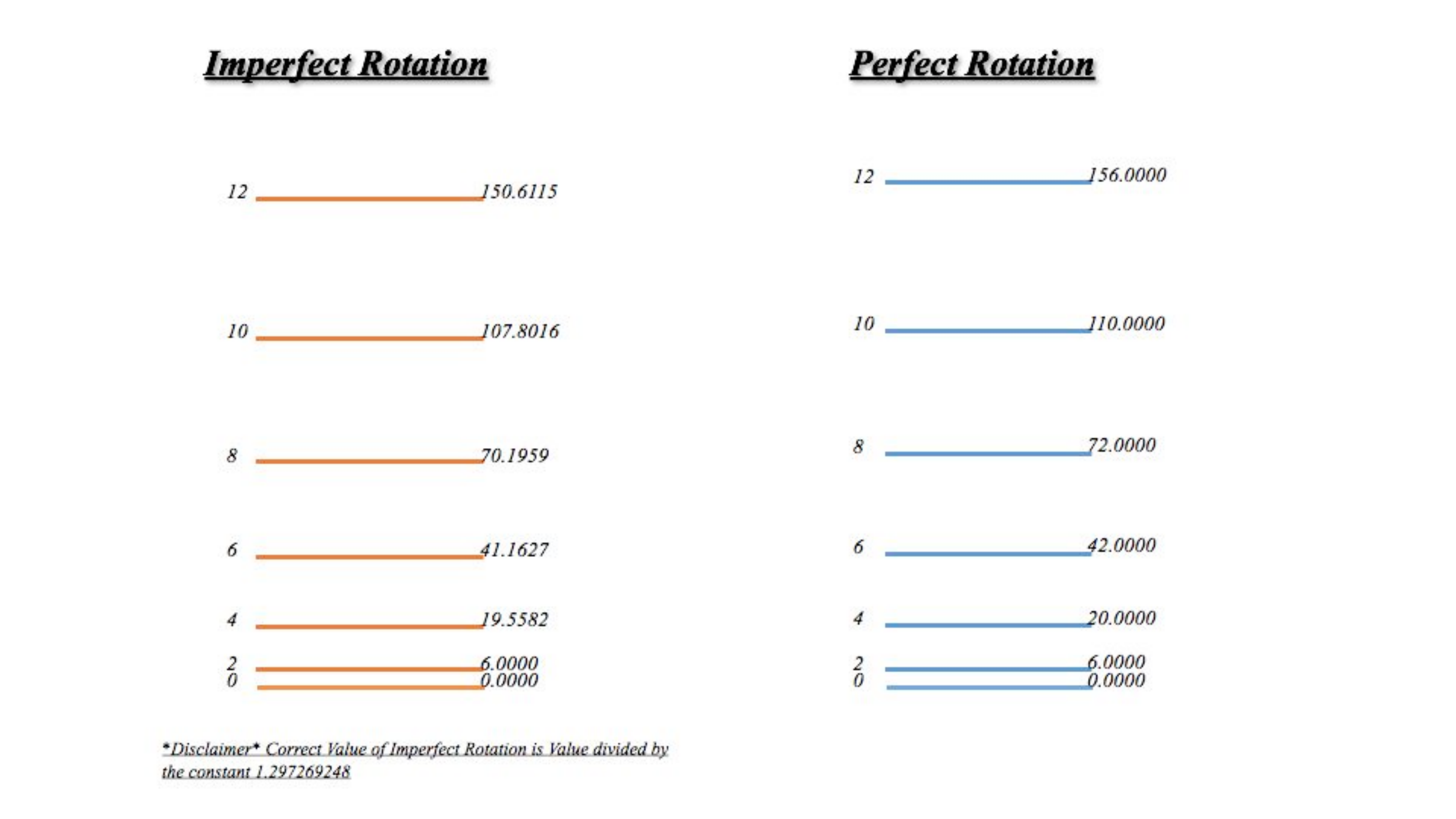}
\end{center}

\section{Fun Spectra}

We here look around for interesting relations between input 2-particle
spectra and more complicated systems. As before, the latter will be
$^{44}$Ti as 2 protons and 2 neutrons in the f$_{7/2}$ shell. We
use trial and error. We have found one interesting case which we here
mention. The two-body interacton matrix elements are 0, 0, 1, 0, 2,
0, 3, 0. We call this the 123 interaction. In other words, all the
T=0 matrix elements are set to zero and the T=1's are set to J/2.
The resulting spectra are shown in table IV. Of special interest is
that the levels from I=6 to I=12 equally spaced with a separation
of 1.5 MeV, almost twice the 2 $\rightarrow$ 0 splitting. Thus, as
shown in table IV, we have achieved a ``vibrational spectrum'' from
I=6 to I=12, but not from I=0 to I=6.

Of even more interest are the structures of the I=6 and I=8 wave functions
in tables VI and VII. That the (2,4) and (4,2) configurations have
the same value is not a surprise. It follows from charge symmetry,
and the same sign from the fact that we are dealing with a T=0 state.
What is a surprise is that for I=6 we have a multitude of zeros --
(2,6), (4,4), (6,2), (6,4) and (6,6). There are corresponding zeros
for I=8. We note one common feature -- things seem to separate into
classes such that configuratons with the same value of the sum of
the proton-proton and neutron-neutron angular momenta act the same.
For I=6 the (6,0) and (4,2) coefficients are non-zero. The J$_{p}$+J$_{n}$
sum is 6. For (2,6) and (4,4) the sum is 8 and for this class the
coefficients are all zero. And (6,6) gives us 12 and here we also
get a zero coefficient.

A more complete picture is afforded if we include the odd angular
momenta. This is shown in Table XXIX. We show all the equally spaced
levels and all the angular momenta of degenerate T=0 states. We note
that for a given energy, all the states have the same value of (J$_{p}$+
J$_{n}).$ That quantity is listed in the first column. Next comes
the energy E, a bit rounded off to show more clearly that the spacing
is 1.5 MeV. Then we list, for a given energy, the angular momenta
of the states with that energy. For example, for 4.65 MeV, all states
have J$_{p}$+ J$_{n}$ = 8. There are 3 degenerate states in this
case with I=6, 7, 8.

If we go to the g$_{9/2}$ shell we get a similar behaviour (now with
the 1234 interaction), but starting from I=8 and ending at I=14. The
spacings are still 1.5 MeV and the I=8 and beyond wave functions are
very strange looking.

Some of the results have been known before. In the f$_{7/2}$ case
there are several states at the same energy 6.15 MeV. They have angular
momenta 3, 7, 9 and 10. But this is not specific to the 123 interaction
-- it applies to any interaction acting only for T=1. This has been
noted and explained {[}10,11,12,13{]} as an example of a partial dynamical
symmetry. They noted that all these T=0 states have the same dual
quantum numbers (J$_{p}$, J$_{n}$), not just the sum (J$_{p}$ +J$_{n}$).
A common feature of these angular momenta is that they cannot occur
or a system of 4 identical nucleons in the f$_{7/2}$ shell e.g. $^{44}$Ca.
This can be seen in the tables of identical-particle fractional parentage
coefficients, B.F. Bayman and A. Lande {[}14{]}.The conditions that
are imposed by the non-existence of these states can be used to show
that indeed the states 3, 7, 9, and 10 with the same (J$_{p}$, J$_{n}$)
are degenerate. What is new in this work is that with the specific
123 (T=1) interaction, two other angular momenta I=6 and 8 enter the
game. Note that we are considering only T=0 states. There are no T=0
states with I=1 or 11 in the (f$_{7/2}$)$^{4}$ configuration. 

We also show more briefly results in table XXX for the 135 interaction
0, 0, 1, 0, 3, 0, 5, 0. The spacings are now 3 MeV, double those for
the 123 interaction. We note that the results for (J$_{p}$+J$_{n})$
= 10 and 12 are the same as in table XXIX. This is to be expected,
since they involve angular momenta not present for 4 identical nucleons
in the f$_{7/2}$ shell and are therefore true for any T=1 interaction.
The main difference is that with 135 there is only one special I=6$^{+}$
state, whereas with 123 there were 2 special 6$^{+}$ states.

If we go to the g$_{9/2}$ shell we get a similar behavour, (now with
the 1234 interaction), but starting from I=7 and ending at I=14. The
results are shown in table XXXI. The spacings are still 1.5 MeV and
the I=7 and beyond wave functions have fixed (J$_{p}$+J$_{n}$).
As an example there are degerate states at 7.29 MeV with I=10, 11,
and 12, as well as states at 8.79 MeV with I=11, 13, and 14. In the
(g$_{9/2}$)$^{4}$ configuration of identical particles, e.g. neutrons,
the following angular momenta cannot occur: 1, 11, 13, 14, 15, 16.
There are no T=0 states with I=1, so this angular momentum is not
under consideration.

In the next section we will be showing wave funcitons as column vectors
with ammplitudes D$^{I}$(J$_{p}$,J$_{n}$), such that for a i state
of tota angular momentum I |D$^{I}$(J$_{p}$,J$_{n}$)|$^{2}$ is
the probaility that the protons couple to J$_{p}$ and the neutrons
to J$_{n}$. Note that for an N=Z nucleus e.g. $^{44}$ Ti one has
the following relation for a state of isospin T:

D(J$_{p}$,J$_{n}$)= (-1)$^{(I+T)}$ D(J$_{n}$,J$_{p}$).

\subsection{123 Tables in f$_{7/2}$ $^{44}$Ti}

\begin{center}
\textbf{Table IV: Fun Spectra and Differences} 
\par\end{center}

\begin{center}
\begin{tabular}{|c|c|c|}
\hline 
I  & E  & Diff.\tabularnewline
\hline 
\hline 
0  & 0.0000  & \tabularnewline
\hline 
2  & 0.7552  & 0.7552\tabularnewline
\hline 
4  & 1.8338  & 1.0786\tabularnewline
\hline 
6  & 3.1498  & 1.3160\tabularnewline
\hline 
8  & 4.6498  & 1.5000\tabularnewline
\hline 
10  & 6.1498  & 1.5000\tabularnewline
\hline 
12  & 7.6498  & 1.5000\tabularnewline
\hline 
\end{tabular}
\par\end{center}

\begin{center}
\textbf{Table V: Wave functions for I=3}
\par\end{center}

\begin{center}
\begin{tabular}{|c|c|c|}
\hline 
J$_{p}$  & J$_{n}$  & D(J$_{p}$,J$_{n}$)\tabularnewline
\hline 
\hline 
2 & 2 & 0.0000\tabularnewline
\hline 
2 & 4 & 0.0000\tabularnewline
\hline 
4 & 2 & 0.0000\tabularnewline
\hline 
4 & 4 & 0.0000\tabularnewline
\hline 
4 & 6 & -0.7071\tabularnewline
\hline 
6 & 4 & 0.7071\tabularnewline
\hline 
6 & 6 & 0.0000\tabularnewline
\hline 
\end{tabular}
\par\end{center}

\begin{center}
\textbf{Table VI: Wave functions for I=6} 
\par\end{center}

\begin{center}
\begin{tabular}{|c|c|c|c|}
\hline 
J$_{p}$  & J$_{n}$  & D(J$_{p}$,J$_{n}$) & D(J$_{p}$,J$_{n}$)\tabularnewline
\hline 
\hline 
0 & 6 & 0.3953 & 0.0000\tabularnewline
\hline 
2 & 4 & 0.5863 & 0.0000\tabularnewline
\hline 
2 & 6 & 0.0000 & -0.4743\tabularnewline
\hline 
4 & 2 & 0.5863 & 0.0000\tabularnewline
\hline 
4 & 4 & 0.0000 & 0.7416\tabularnewline
\hline 
4 & 6 & 0.0000 & 0.0000\tabularnewline
\hline 
6 & 0 & 0.3953 & 0.0000\tabularnewline
\hline 
6 & 2 & 0.0000 & -0.4743\tabularnewline
\hline 
6 & 4 & 0.0000 & 0.0000\tabularnewline
\hline 
6 & 6 & 0.0000 & 0.0000\tabularnewline
\hline 
\end{tabular}
\par\end{center}

\begin{center}
\textbf{Table VII: Wave functions for I=8} 
\par\end{center}

\begin{center}
\begin{tabular}{|c|c|c|}
\hline 
J$_{p}$  & J$_{n}$  & D(J$_{p}$,J$_{n}$)\tabularnewline
\hline 
\hline 
2  & 6  & 0.4882\tabularnewline
\hline 
4  & 4  & 0.7234\tabularnewline
\hline 
4  & 6  & 0.0000\tabularnewline
\hline 
6  & 2  & 0.4882\tabularnewline
\hline 
6  & 4  & 0.0000\tabularnewline
\hline 
6  & 6  & 0.0000\tabularnewline
\hline 
\end{tabular}
\par\end{center}

\begin{center}
\textbf{Table VIII: Wave functions for I=9}
\par\end{center}

\begin{center}
\begin{tabular}{|c|c|c|}
\hline 
J$_{p}$  & J$_{n}$  & D(J$_{p}$,J$_{n}$)\tabularnewline
\hline 
\hline 
4 & 6 & -0.7071\tabularnewline
\hline 
6 & 4 & 0.7071\tabularnewline
\hline 
6 & 6 & 0.0000\tabularnewline
\hline 
\end{tabular}
\par\end{center}

\begin{center}
\textbf{Table IX: Wave functions for I=10}
\par\end{center}

\begin{center}
\begin{tabular}{|c|c|c|}
\hline 
J$_{p}$  & J$_{n}$  & D(J$_{p}$,J$_{n}$)\tabularnewline
\hline 
\hline 
4 & 6 & 0.7071\tabularnewline
\hline 
6 & 4 & 0.7071\tabularnewline
\hline 
6 & 6 & 0.0000\tabularnewline
\hline 
\end{tabular}
\par\end{center}

\begin{center}
\textbf{Table X: Wave functions for I=12}
\par\end{center}

\begin{center}
\begin{tabular}{|c|c|c|}
\hline 
J$_{p}$  & J$_{n}$  & D(J$_{p}$,J$_{n}$)\tabularnewline
\hline 
\hline 
6 & 6 & 1.0000\tabularnewline
\hline 
\end{tabular}
\par\end{center}

\subsection{135 Tables in f$_{7/2}$}

\begin{center}
\textbf{Table XI: Wave functions for I=3}
\par\end{center}

\begin{center}
\begin{tabular}{|c|c|c|c|}
\hline 
J$_{p}$  & J$_{n}$  & D(J$_{p}$,J$_{n}$) & D(J$_{p}$,J$_{n}$)\tabularnewline
\hline 
\hline 
2 & 2 & 0.0000 & 0.0000\tabularnewline
\hline 
2 & 4 & -0.7071 & 0.0000\tabularnewline
\hline 
4 & 2 & 0.7071 & 0.0000\tabularnewline
\hline 
4 & 4 & 0.0000 & 0.0000\tabularnewline
\hline 
4 & 6 & 0.0000 & -0.7071\tabularnewline
\hline 
6 & 4 & 0.0000 & 0.7071\tabularnewline
\hline 
6 & 6 & 0.0000 & 0.0000\tabularnewline
\hline 
\end{tabular}
\par\end{center}

\begin{center}
\textbf{Table XII: Wave functions for I=6}
\par\end{center}

\begin{center}
\begin{tabular}{|c|c|c|}
\hline 
J$_{p}$  & J$_{n}$  & D(J$_{p}$,J$_{n}$)\tabularnewline
\hline 
\hline 
0 & 6 & 0.0000\tabularnewline
\hline 
2 & 4 & 0.0000\tabularnewline
\hline 
2 & 6 & -0.4743\tabularnewline
\hline 
4 & 2 & 0.0000\tabularnewline
\hline 
4 & 4 & 0.7416\tabularnewline
\hline 
4 & 6 & 0.0000\tabularnewline
\hline 
6 & 0 & 0.0000\tabularnewline
\hline 
6 & 2 & -0.4743\tabularnewline
\hline 
6 & 4 & 0.0000\tabularnewline
\hline 
6 & 6 & 0.0000\tabularnewline
\hline 
\end{tabular}
\par\end{center}

\begin{center}
\textbf{Table XIII: Wave functions for I=7}
\par\end{center}

\begin{center}
\begin{tabular}{|c|c|c|}
\hline 
J$_{p}$  & J$_{n}$  & D(J$_{p}$,J$_{n}$)\tabularnewline
\hline 
\hline 
2 & 6 & 0.0000\tabularnewline
\hline 
4 & 4 & 0.0000\tabularnewline
\hline 
4 & 6 & -0.7071\tabularnewline
\hline 
6 & 2 & 0.0000\tabularnewline
\hline 
6 & 4 & 0.7071\tabularnewline
\hline 
6 & 6 & 0.0000\tabularnewline
\hline 
\end{tabular}
\par\end{center}

\begin{center}
\textbf{Table XIV: Wave functions for I=8}
\par\end{center}

\begin{center}
\begin{tabular}{|c|c|c|}
\hline 
J$_{p}$  & J$_{n}$  & D(J$_{p}$,J$_{n}$)\tabularnewline
\hline 
\hline 
2 & 6 & 0.4882\tabularnewline
\hline 
4 & 4 & 0.7234\tabularnewline
\hline 
4 & 6 & 0.0000\tabularnewline
\hline 
6 & 2 & 0.4882\tabularnewline
\hline 
6 & 4 & 0.0000\tabularnewline
\hline 
6 & 6 & 0.0000\tabularnewline
\hline 
\end{tabular}
\par\end{center}

\begin{center}
\textbf{Table XV: Wave functions for I=9}
\par\end{center}

\begin{center}
\begin{tabular}{|c|c|c|}
\hline 
J$_{p}$  & J$_{n}$  & D(J$_{p}$,J$_{n}$)\tabularnewline
\hline 
\hline 
4 & 6 & -0.7071\tabularnewline
\hline 
6 & 4 & 0.7071\tabularnewline
\hline 
6 & 6 & 0.0000\tabularnewline
\hline 
\end{tabular}
\par\end{center}

\begin{center}
\textbf{Table XVI: Wave functions for I=10}
\par\end{center}

\begin{center}
\begin{tabular}{|c|c|c|}
\hline 
J$_{p}$  & J$_{n}$  & D(J$_{p}$,J$_{n}$)\tabularnewline
\hline 
\hline 
4 & 6 & 0.7071\tabularnewline
\hline 
6 & 4 & 0.7071\tabularnewline
\hline 
6 & 6 & 0.0000\tabularnewline
\hline 
\end{tabular}
\par\end{center}

\begin{center}
\textbf{Table XVII: Wave functions for I=12}
\par\end{center}

\begin{center}
\begin{tabular}{|c|c|c|}
\hline 
J$_{p}$  & J$_{n}$  & D(J$_{p}$,J$_{n}$)\tabularnewline
\hline 
\hline 
6 & 6 & 1.0000\tabularnewline
\hline 
\end{tabular}
\par\end{center}

\subsection{1234 Tables in g$_{9/2}$$^{96}$Cd}

\begin{center}
\textbf{Table XVIII: Wave functions for I=3}
\par\end{center}

\begin{center}
\begin{tabular}{|c|c|c|}
\hline 
J$_{p}$  & J$_{n}$  & D(J$_{p}$,J$_{n}$)\tabularnewline
\hline 
\hline 
2 & 2 & 0.0000\tabularnewline
\hline 
2 & 4 & -0.6598\tabularnewline
\hline 
4 & 2 & 0.6598\tabularnewline
\hline 
4 & 4 & 0.0000\tabularnewline
\hline 
4 & 6 & -0.2449\tabularnewline
\hline 
6 & 4 & 0.2449\tabularnewline
\hline 
6 & 6 & 0.0000\tabularnewline
\hline 
6 & 8 & 0.0690\tabularnewline
\hline 
8 & 6 & -0.0690\tabularnewline
\hline 
8 & 8 & 0.0000\tabularnewline
\hline 
\end{tabular}
\par\end{center}

\begin{center}
\textbf{Table XIX: Wave functions for I=7}
\par\end{center}

\begin{center}
\begin{tabular}{|c|c|c|}
\hline 
J$_{p}$  & J$_{n}$  & D(J$_{p}$,J$_{n}$)\tabularnewline
\hline 
\hline 
2 & 6 & 0.0000\tabularnewline
\hline 
2 & 8 & -0.3459\tabularnewline
\hline 
4 & 4 & 0.0000\tabularnewline
\hline 
4 & 6 & 0.6167\tabularnewline
\hline 
4 & 8 & 0.0000\tabularnewline
\hline 
6 & 2 & 0.0000\tabularnewline
\hline 
6 & 4 & -0.6167\tabularnewline
\hline 
6 & 6 & 0.0000\tabularnewline
\hline 
6 & 8 & 0.0000\tabularnewline
\hline 
8 & 2 & 0.3459\tabularnewline
\hline 
8 & 4 & 0.0000\tabularnewline
\hline 
8 & 6 & 0.0000\tabularnewline
\hline 
8 & 8 & 0.0000\tabularnewline
\hline 
\end{tabular}
\par\end{center}

\begin{center}
\textbf{Table XX: Wave functions for I=8}
\par\end{center}

\begin{center}
\begin{tabular}{|c|c|c|c|}
\hline 
J$_{p}$  & J$_{n}$  & D(J$_{p}$,J$_{n}$) & D(J$_{p}$,J$_{n}$)\tabularnewline
\hline 
\hline 
0 & 8 & 0.2792 & -0.2173\tabularnewline
\hline 
2 & 6 & 0.4949 & 0.1322\tabularnewline
\hline 
2 & 8 & 0.0000 & 0.4214\tabularnewline
\hline 
4 & 4 & 0.5951 & 0.0000\tabularnewline
\hline 
4 & 6 & 0.0000 & 0.0000\tabularnewline
\hline 
4 & 8 & 0.0000 & -0.4985\tabularnewline
\hline 
6 & 2 & 0.4949 & -0.1322\tabularnewline
\hline 
6 & 4 & 0.0000 & 0.0000\tabularnewline
\hline 
6 & 6 & 0.0000 & 0.0000\tabularnewline
\hline 
6 & 8 & 0.0000 & 0.0962\tabularnewline
\hline 
8 & 0 & 0.2792 & 0.2173\tabularnewline
\hline 
8 & 2 & 0.0000 & -0.4214\tabularnewline
\hline 
8 & 4 & 0.0000 & 0.4985\tabularnewline
\hline 
8 & 6 & 0.0000 & -0.0962\tabularnewline
\hline 
8 & 8 & 0.0000 & 0.0000\tabularnewline
\hline 
\end{tabular}
\par\end{center}

\begin{center}
\textbf{Table XXI: Wave functions for I=9}
\par\end{center}

\begin{center}
\begin{tabular}{|c|c|c|}
\hline 
J$_{p}$  & J$_{n}$  & D(J$_{p}$,J$_{n}$)\tabularnewline
\hline 
\hline 
2 & 8 & 0.6281\tabularnewline
\hline 
4 & 6 & 0.3248\tabularnewline
\hline 
4 & 8 & 0.0000\tabularnewline
\hline 
6 & 4 & -0.3248\tabularnewline
\hline 
6 & 6 & 0.0000\tabularnewline
\hline 
6 & 8 & 0.0000\tabularnewline
\hline 
8 & 2 & -0.6281\tabularnewline
\hline 
8 & 4 & 0.0000\tabularnewline
\hline 
8 & 6 & 0.0000\tabularnewline
\hline 
8 & 8 & 0.0000\tabularnewline
\hline 
\end{tabular}
\par\end{center}

\begin{center}
\textbf{Table XXII: Wave functions for I=10}
\par\end{center}

\begin{center}
\begin{tabular}{|c|c|c|c|}
\hline 
J$_{p}$  & J$_{n}$  & D(J$_{p}$,J$_{n}$) & D(J$_{p}$,J$_{n}$)\tabularnewline
\hline 
\hline 
2 & 8 & 0.3293 & 0.0000\tabularnewline
\hline 
4 & 6 & 0.6258 & 0.0000\tabularnewline
\hline 
4 & 8 & 0.0000 & -0.5126\tabularnewline
\hline 
6 & 4 & 0.6258 & 0.0000\tabularnewline
\hline 
6 & 6 & 0.0000 & 0.6888\tabularnewline
\hline 
6 & 8 & 0.0000 & 0.0000\tabularnewline
\hline 
8 & 2 & 0.3293 & 0.0000\tabularnewline
\hline 
8 & 4 & 0.0000 & -0.5126\tabularnewline
\hline 
8 & 6 & 0.0000 & 0.0000\tabularnewline
\hline 
8 & 8 & 0.0000 & 0.0000\tabularnewline
\hline 
\end{tabular}
\par\end{center}

\begin{center}
\textbf{Table XXIII: Wave functions for I=11}
\par\end{center}

\begin{center}
\begin{tabular}{|c|c|c|c|}
\hline 
J$_{p}$  & J$_{n}$  & D(J$_{p}$,J$_{n}$) & D(J$_{p}$,J$_{n}$)\tabularnewline
\hline 
\hline 
4 & 8 & 0.7071 & 0.0000\tabularnewline
\hline 
6 & 6 & 0.0000 & 0.0000\tabularnewline
\hline 
6 & 8 & 0.0000 & 0.7071\tabularnewline
\hline 
8 & 4 & -0.7071 & 0.0000\tabularnewline
\hline 
8 & 6 & 0.0000 & -0.7071\tabularnewline
\hline 
8 & 8 & 0.0000 & 0.0000\tabularnewline
\hline 
\end{tabular}
\par\end{center}

\begin{center}
\textbf{Table XXIV: Wave functions for I=12}
\par\end{center}

\begin{center}
\begin{tabular}{|c|c|c|}
\hline 
J$_{p}$  & J$_{n}$  & D(J$_{p}$,J$_{n}$)\tabularnewline
\hline 
\hline 
4 & 8 & 0.4715\tabularnewline
\hline 
6 & 6 & 0.7452\tabularnewline
\hline 
6 & 8 & 0.0000\tabularnewline
\hline 
8 & 4 & 0.4715\tabularnewline
\hline 
8 & 6 & 0.0000\tabularnewline
\hline 
8 & 8 & 0.0000\tabularnewline
\hline 
\end{tabular}
\par\end{center}

\begin{center}
\textbf{Table XXV: Wave functions for I=13}
\par\end{center}

\begin{center}
\begin{tabular}{|c|c|c|}
\hline 
J$_{p}$  & J$_{n}$  & D(J$_{p}$,J$_{n}$)\tabularnewline
\hline 
\hline 
6 & 8 & 0.7071\tabularnewline
\hline 
8 & 6 & -0.7071\tabularnewline
\hline 
8 & 8 & 0.0000\tabularnewline
\hline 
\end{tabular}
\par\end{center}

\begin{center}
\textbf{Table XXVI: Wave functions for I=14}
\par\end{center}

\begin{center}
\begin{tabular}{|c|c|c|}
\hline 
J$_{p}$  & J$_{n}$  & D(J$_{p}$,J$_{n}$)\tabularnewline
\hline 
\hline 
6 & 8 & 0.7071\tabularnewline
\hline 
8 & 6 & 0.7071\tabularnewline
\hline 
8 & 8 & 0.0000\tabularnewline
\hline 
\end{tabular}
\par\end{center}

\begin{center}
\textbf{Table XXVII: Wave functions for I=15}
\par\end{center}

\begin{center}
\begin{tabular}{|c|c|c|}
\hline 
J$_{p}$  & J$_{n}$  & D(J$_{p}$,J$_{n}$)\tabularnewline
\hline 
\hline 
8 & 8 & 1.0000\tabularnewline
\hline 
\end{tabular}
\par\end{center}

\begin{center}
\textbf{Table XXVIII: Wave functions for I=16}
\par\end{center}

\begin{center}
\begin{tabular}{|c|c|c|}
\hline 
J$_{p}$  & J$_{n}$  & D(J$_{p}$,J$_{n}$)\tabularnewline
\hline 
\hline 
8 & 8 & 1.0000\tabularnewline
\hline 
\end{tabular}
\par\end{center}

\begin{center} 
  \includegraphics[width=.75\textwidth]{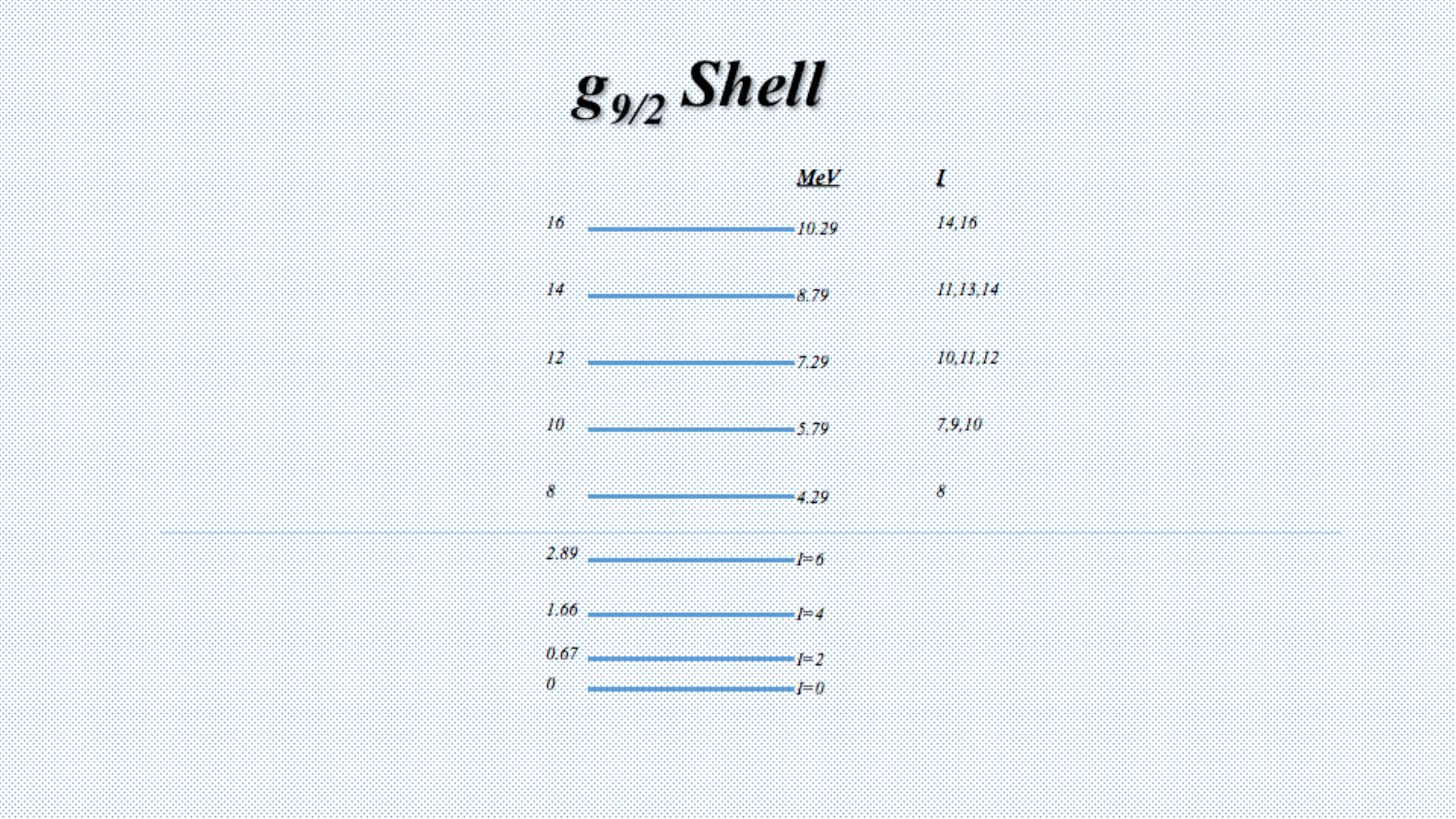}
\end{center}

\subsection{Two protons and two neutrons in the h$_{11/2}$ with the 12345 interaction}

In this section we will be more brief. We will just show in table
XXXII the special states with (J$_{p}$+J$_{n})$ constant for 2 protons
and 2 neutrons in the h$_{11/2}$ shell. We use the 12345 interaction
0, 0, 1, 0, 2, 0, 3, 0, 4, 0, 5, 0. The angular momenta that cannot
occur in this case are 1, 15, 17, 18, 19, and 20. Since we are considering
only T=0 states, we do not consider I=1 or I=19.

\begin{center}
\textbf{Table XXIX: Special states in the f$_{7/2}$shell (123 interaction)}
\par\end{center}

\begin{center}
\begin{tabular}{|c|c|c|}
\hline 
J$_{p}$+J$_{n}$ & E (MeV)  & I\tabularnewline
\hline 
\hline 
6  & 3.15  & 3, 6\tabularnewline
\hline 
8  & 4.65  & 6, 7, 8\tabularnewline
\hline 
10  & 6.15  & 3, 7, 9, 10\tabularnewline
\hline 
12  & 7.65  & 10, 12\tabularnewline
\hline 
\end{tabular}
\par\end{center}

\begin{center}
\textbf{Table XXX: Special states in the f$_{7/2}$shell (135 interaction)}
\par\end{center}

\begin{center}
\begin{tabular}{|c|c|c|}
\hline 
J$_{p}$+ J$_{n}$  & E(MeV)  & I\tabularnewline
\hline 
6  & 4.42  & 3\tabularnewline
\hline 
8  & 7.42  & 6, 7, 8\tabularnewline
\hline 
10  & 10.42  & 3, 7, 9, 10\tabularnewline
\hline 
12  & 13.42  & 10, 12\tabularnewline
\hline 
\end{tabular}
\par\end{center}

\begin{center}
\textbf{Table XXXI: Special states in the g$_{9/2}$shell (1234 interaction)}
\par\end{center}

\begin{center}
\begin{tabular}{|c|c|c|}
\hline 
J$_{p}$+J$_{n}$ & E(MeV) & I\tabularnewline
\hline 
8 & 4.29 & 8\tabularnewline
\hline 
10 & 5.79 & 7, 9,10\tabularnewline
\hline 
12 & 7.29 & 10, 11, 12\tabularnewline
\hline 
14 & 8.79 & 11, 13, 14\tabularnewline
\hline 
16 & 10.29 & 14, 16\tabularnewline
\hline 
\end{tabular}
\par\end{center}

\begin{center}
\textbf{Table: XXXII: Special states in the h$_{11/2}$shell (12345
interaction)}
\par\end{center}

\begin{center}
\begin{tabular}{|c|c|c|}
\hline 
J$_{p}$+J$_{n}$ & E (MeV) & I\tabularnewline
\hline 
\hline 
10 & 5.46 & 10\tabularnewline
\hline 
12 & 6.96 & 11, 12\tabularnewline
\hline 
14 & 8.46 & 11, 13, 14\tabularnewline
\hline 
16 & 9.96 & 14, 15, 16\tabularnewline
\hline 
18 & 11.46 & 15, 17, 18\tabularnewline
\hline 
20 & 12.96 & 18, 20\tabularnewline
\hline 
\end{tabular}
\par\end{center}

\begin{center} 
  \includegraphics[width=.75\textwidth]{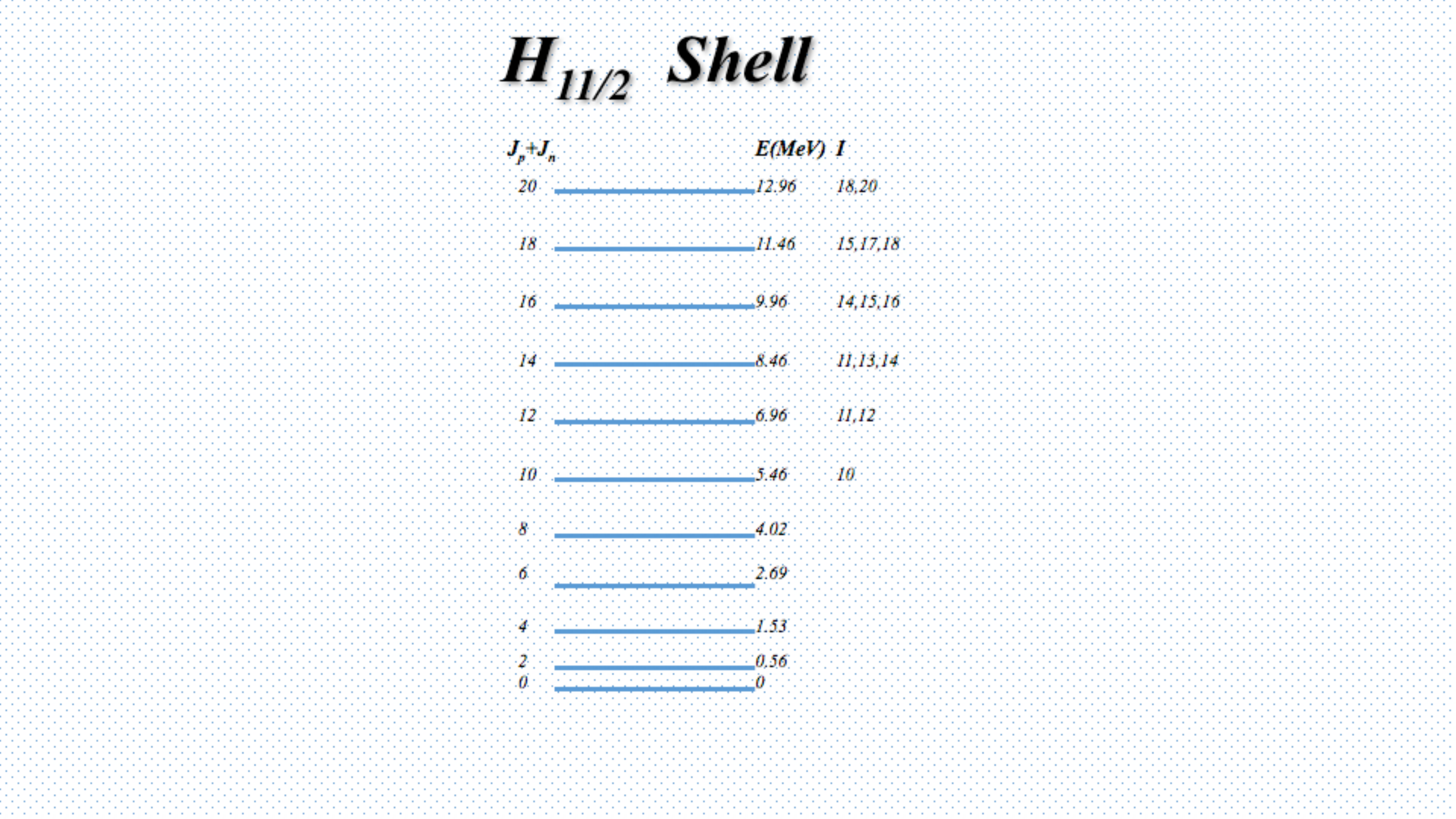}
\end{center}

\subsection{Explanation} 
  For a detailed explanation of hte results we refer the reader to a final version of this work
W.Pereira, R. Garcia, L. Zamick, A,Escuderos and K. Neergaard, Int .J.  Mod. Phys. E26,1740021 (2017)

\begin{table}

\begin{center}

\textbf{Table: XXXIII: Special states not shown in previous tables}

\bigskip

\begin{tabular}{|c|c|c|}
\hline 
$j$ & $J_p + J_n$ & $I$\\
\hline 
\hline 
$1/2$ & 0 & 0\\
\hline 
$3/2$ & 2 & 2\\
\hline 
& 4 & 2, 4\\
\hline 
$5/2$ & 4 & 4\\
\hline 
& 6 & 3, 5, 6\\
\hline 
& 8 & 6, 8\\
\hline 
13/2 & 12 & 12\\
\hline 
& 14 & 13, 14\\
\hline 
& 16 & 15, 16\\
\hline 
& 18 & 15, 17, 18\\
\hline 
& 20 & 18, 19, 20\\
\hline 
& 22 & 19, 21, 22\\
\hline 
& 24 & 22, 24\\
\hline 
15/2 & 14 & 14\\
\hline 
& 16 & 15, 16\\
\hline 
& 18 & 17, 18\\
\hline 
& 20 & 19, 20\\
\hline 
& 22 & 19, 21, 22\\
\hline 
& 24 & 22, 23, 24\\
\hline 
& 26 & 23, 25, 26\\
\hline 
& 28 & 26, 28\\
\hline
\end{tabular}

\end{center}

\end{table}

\subsection{Final Comment}

Preiviously we had found a partial dynamical symmetry when we set
all T=0 two-body amtrix elemnts to zero in a single j shell calculation
for 2 protons and 2 neutrons {[}10.11.1,12,13{]}. The angular momenta
involved were those that could not occur for 4 identical particles.
For the states in question one had (J$_{p}$, J$_{n}$) as good dual
quantum numbers. When we consider more restricitve interactions, still
with T=0 interactions set to zero e.g. ``123'' in f$_{7/2}$, ``1234''
in g$_{9/2}$ ,''12345'' in h$_{11/2}$ we get some selected equally
spaced levels which are usually multi-degenerate. As well as the old
we get new angular momenta as part of the eaually spaced spectra.These
can occur for systems of identical particles. The wave functions have
the constraint that ( J$_{p}$ + J$_{n}$) is a constant. This is
less constrictive than the previous condition.

\subsection{Acknowledgments}

Wesley Pereira is a student at Essex College, Newark, New Jersey,
07102. His research at Rutgers is funded by a Garden State Stokes
Alliance for Minorities Participation (G.S.L.S.A.M.P.) internship.

Ricardo Garcia has two institutional affiliations: Rutgers University,
and the University of Puerto Rico, Rio Piedras Campus. The permament
address associated with the UPR-RP is University of Puerto Rico, San
Juan, Puerto Rico 00931. He acknowledges that to carry out this work
he has received support via the Research Undergraduate Experience
program (REU) from the U.S. National Science Foundation through grant
PHY-1263280, and thanks the REU Physics program at Rutgers University
for their support.

\end{document}